# Charged particle in the field an electric quadrupole in two dimensions

**A. D. Alhaidari**

*Shura Council, Riyadh 11212, Saudi Arabia*
AND
*Physics Department, King Fahd University of Petroleum and Minerals, Dhahran 31261, Saudi Arabia*

email: haidari@mailaps.org

We obtain analytic solution of the time-independent Schrödinger equation in two dimensions for a charged particle moving in the field of an electric quadrupole. The solution is written as a series in terms of special functions that support a tridiagonal matrix representation for the angular and radial components of the wave operator. This solution is for all energies, the discrete (for bound states) as well as the continuous (for scattering states). The expansion coefficients of the wavefunction are written in terms of orthogonal polynomials satisfying three-term recursion relations. The charged particle could become bound to the quadrupole only if its moment exceeds a certain critical value.

## 1. Introduction

The interaction of a charged particle with an electric dipole and the formation of bounded anions is a fundamental problem, which received a lot of attention in the physics literature for more than 60 years [1,2]. This interest was fueled by the observation that electron capture by a molecule with a permanent electric dipole moment could take place only if the dipole moment exceeds a certain minimum value which does not depend on the size of the dipole [2]. Moreover, the electron binding properties of polar molecules have been the topic of considerable theoretical and experimental interest [3]. Despite renewed interest in this problem, no exact analytic solution (aside from our recent contribution [4]) was reported in the literature. This is because, even in the ideal case of a point electric dipole in 3D, the interaction is represented by the intractable non-central potential $\cos\theta/r^2$ (in spherical coordinates) which was known not to belong to any of the well-established classes of exactly solvable problems. However, our approach [4] followed another path where we employ the tools of our "Tridiagonal Physics" program [5]. In this program, which is inspired by the J-matrix method [6], we relax the usual restriction of a diagonal representation on the solution space of the eigenvalue wave equation. We only require that the matrix representation of the wave operator (radial and angular) be tridiagonal and symmetric. Consequently, this reduced constraint results in a solution space that is large enough to incorporate an exact analytic solution for this noncentral potential. In this paper, we extend those developments to the electric quadrupole in two dimensions. This constitutes the first contribution of its kind to the solution of this problem.

Now, the Coulomb potential of a point charge (electric monopole) in two dimensions is logarithmic in the distance from the charge. In Appendix A, we address the issue of the Coulomb problem in any dimension and calculate the potentials of a point electric dipole and quadrupole in two dimensions. The problem of a point charge in the field of an electric quadrupole in 2D (Fig. 2) is *equivalent* to the problem of a point charge in 3D moving in a plane normal to four parallel uniform line charges whose configuration is shown in Fig. 4. Consequently, the findings in this work (e.g., existence of a critical quadrupole moment for electron capture) could, in fact, be tested in an appropriate experimental setup.

The two-dimensional time-independent Schrödinger equation for a particle of mass $m$ and charge $q$ in an electrostatic potential $V(\vec{r})$ is

$$\left[-\tfrac{1}{2}\vec{\nabla}^2 + V(\vec{r}) - E\right]\psi = 0, \tag{1.1}$$

where $\vec{\nabla}$ is the two-dimensional Laplacian and we have used the atomic units $\hbar = m = q = 1$. The energy, $E$, is real and it is either discrete for bound states, or continuous for scattering states. In the 2D cylindrical coordinates, $\vec{r} = \{r, \theta\}$, this wave equation could be written explicitly as follows

$$\left\{\frac{1}{r}\frac{\partial}{\partial r} r \frac{\partial}{\partial r} + \frac{4}{r^2}\left[(1-x^2)\frac{\partial^2}{\partial x^2} - x\frac{\partial}{\partial x}\right] - 2V + 2E\right\}\psi = 0, \tag{1.2}$$

where $x = \sin 2\theta$ chosen to be compatible with the angular dependence of the quadrupole potential given in Appendix A by Eq. (A.4). This choice of variables results in a one-to-one map between $x$ and $\theta$ for only a quarter of the 2D plane (e.g., $\theta \in \left[-\tfrac{\pi}{4}, +\tfrac{\pi}{4}\right]$). Symmetry of the problem (see, for example, Fig. 2) shows that for the complete solution of the problem it is sufficient to obtain a solution in only a quarter of the plane. Now, Eq. (1.2) is separable for potentials of the form

$$V(\vec{r}) = V_r(r) + \frac{1}{r^2} V_\theta(x). \tag{1.3}$$

This is so because if we write the wavefunction as $\psi(r,\theta) = r^{-\tfrac{1}{2}} R(r) \Theta(\theta)$, then the wave equation (1.2) with the potential (1.3) becomes separated in the two coordinates as follows

$$\left[(1-x^2)\frac{d^2}{dx^2} - x\frac{d}{dx} - \frac{1}{2}V_\theta + \frac{1}{2}E_\theta\right]\Theta = 0, \tag{1.4a}$$

$$\left(\frac{d^2}{dr^2} - \frac{2E_\theta}{r^2} - 2V_r + 2E\right)R = 0, \tag{1.4b}$$

where $E_\theta$ is the separation constant, which is real and dimensionless. Square integrability is with respect to the following integration measures

$$\int |\psi|^2 d^2\vec{r} = 2\int_0^\infty |R|^2 dr \int_{-1}^{+1} (1-x^2)^{-\tfrac{1}{2}} |\Theta|^2 dx. \tag{1.5}$$

Moreover, the components of the wave function must satisfy satisfy: $R(0) = R(\infty) = 0$, and $\Theta(\theta) = \Theta(\theta + 2\pi)$.

Now, we consider the problem of a charged particle moving in the field of a two-dimensional electric quadrupole of moment $p$ (see, Fig. 2). Thus, as shown in Appendix A, the potential components become: $V_r = 0$, $V_\theta = -(4p/qa_0^2)\sin 2\theta$, where the length scale $a_0 = 4\pi\epsilon_0 \hbar^2/mq^2$ (for an electron, this is the Bohr radius). If we measure length (e.g., $r$ and $\sqrt{p/q}$) in units of $a_0$ then the energy will be given in units of $m^{-1}(\hbar/a_0)^2$ and $V_\theta \to -4(p/q)\sin 2\theta$. Moreover, the set of equations (1.4a) and (1.4b) show that the system's energy $E$ is a function of $E_\theta$, which in turn is a function of the quadrupole moment. With $V_r = 0$, Eq. (1.4b) becomes the wave equation for the inverse square potential. It has been well established that bound state solutions in this potential could be supported only if the dimensionless coupling parameter $2E_\theta$ is less than the critical value $-\frac{1}{4}$ [7]. Therefore, for bound states we take $2E_\theta = -\omega^2 - \frac{1}{4}$, whereas for the continuum scattering states we write $2E_\theta = \gamma^2 - \frac{1}{4}$, where $\gamma$ and $\omega$ are real dimensionless parameters. In the following two sections, we obtain the angular and radial components of the wave function for both cases. These will be written as series in terms of discrete square integrable functions that support a tridiagonal matrix representation for the wave operator.

## 2. The angular component of the wavefunction

We expand the angular component of the wavefunction, $\Theta(\theta)$, in a complete basis functions $\{\chi_n(x)\}_{n=0}^\infty$ as $\Theta(\theta) = \sum_{n=0}^\infty f_n(E_\theta)\chi_n(x)$. These basis elements must satisfy the boundary conditions in the configuration space with coordinate $x \in [-1,+1]$. We write them as [8]

$$\chi_n(x) = a_n(1-x^2)^\nu D_n^\mu(x), \tag{2.1}$$

where $D_n^\mu(x)$ are the "improved" ultra-spherical Gegenbauer (IUSG) polynomials shown in Appendix B [9] and $n = 0,1,2,...$ The dimensionless real parameters $\nu \geq 0$, $\mu > -\frac{1}{2}$ and $a_n$ is the normalization constant associated with the orthogonality of these polynomials,

$$a_n = \frac{\sqrt{2(n+\mu)}}{2^\mu \Gamma(\mu+1/2)} \sqrt{\frac{\Gamma(n+2\mu)}{\Gamma(n+1)}}. \tag{2.2}$$

Using the differential equation (B.8) and differential formula (B.9) for these IUSG polynomials, shown in Appendix B, we can write the action of the angular differential operator of Eq. (1.4a) on the basis element (2.1) as follows

$$(H_\theta - E_\theta)|\chi_n\rangle = \left\{\frac{-4}{1-x^2}\left[\nu(2\nu-1) + n(2\nu-\mu)\right] + 2(n+2\nu)^2 + V_\theta - E_\theta\right\}|\chi_n\rangle \tag{2.3}$$
$$- 4n(\mu - 2\nu)\frac{x}{1-x^2}\frac{a_n}{a_{n-1}}|\chi_{n-1}\rangle$$

The recurrence relation (B.6) and orthogonality formula (B.10) of the IUSG polynomials show that a tridiagonal matrix representation for $\langle \chi_n | H_\theta - E_\theta | \chi_{n'}\rangle$ is achievable with $V_\theta \sim x$ if and only if $\mu = 2\nu$ and $\mu = 0$ or $1$. The case $\mu = 1$ is to be rejected since the associated ground state wavefunction is not nodeless. Now, to obtain an explicit expression for the matrix representation $\langle \chi_n | H_\theta - E_\theta | \chi_{n'}\rangle$, we project the action of the angular wave operator given by (2.3) on the basis elements from left. Employing again

the orthogonality property and recurrence relation of the IUSG polynomials, we obtain the following tridiagonal structure:

$$\langle \chi_n | H_\theta - E_\theta | \chi_{n'} \rangle = \left(2n^2 - E_\theta\right)\delta_{n,n'} - 2\xi\left(\delta_{n,n'+1} + \delta_{n,n'-1}\right), \qquad (2.4)$$

where $\xi = p/q$. Now, we start by considering the continuum scattering states and write $2E_\theta \equiv \gamma^2 - \frac{1}{4}$, where $\gamma$ is a dimensionless real parameter. Equation (1.4b) shows that $\gamma$ plays the role of the azimuthal quantum number in cylindrically symmetric 2D problems. However, unlike the azimuthal number that assumes only integral values, $\gamma$ is a continuous parameter. The tridiagonal matrix representation of the angular wave operator in (2.4) makes the wave equation (1.4a) equivalent to the following three-term recursion relation for the expansion coefficients of the angular wavefunction

$$\tfrac{1}{4}\gamma^2 f_n^\xi = \left(n^2 + \tfrac{1}{16}\right)f_n^\xi - \xi\left(f_{n-1}^\xi + f_{n+1}^\xi\right). \qquad (2.5)$$

This relation implies that if $\xi$ becomes too large then reality of the representation will be violated. To maintain reality, the quadrupole moment should not exceed a certain critical value. It is the smallest value of $\xi$ that makes the right side of (2.5) vanish. Below, we show how to calculate these critical values. The solution of the recursion relation (2.5) for a fixed $\xi$ is a functions in $\gamma$ that is defined modulo an arbitrary non-singular factor that depends on $\gamma$ but otherwise independent of $n$. If we choose the standard normalization by taking the initial seed as $f_0^\xi(\gamma) = 1$ then the results are polynomials in $\tfrac{1}{4}\gamma^2$. To the best of our knowledge, these polynomials were not studied elsewhere. Since they are associated with the electric quadrupole potential $\xi\frac{\sin 2\theta}{r^2}$ we refer to them as the "quadrupole polynomials". They are completely defined to all degrees by the recursion (2.5) and the initial values $f_0^\xi = 1$ and $f_1^\xi = \tfrac{1}{4\xi}\left(\tfrac{1}{4} - \gamma^2\right)$. Now, since the Chebyshev polynomials of the first kind is related to the IUSG polynomial as $T_n(x) = D_n^0(x)$ then we can write the angular component of the wavefunction for the continuum scattering states as the $L^2$-series

$$\Theta(\theta) = A_\gamma^\xi \sum_{n=0}^\infty \sqrt{(2-\delta_{n0})\pi^{-1}} f_n^\xi(\gamma) T_n(x), \qquad (2.6)$$

where $A_\gamma^\xi$ is a normalization constant that depends on the physical parameters of the problem but, otherwise, independent of $\theta$ and $n$. To make $\Theta(\theta)$ $\gamma$-normalizable, we write $A_\gamma^\xi = \sqrt{\rho^\xi(\gamma)}$, where $\rho^\xi(\gamma)$ is the weight (density) function associated with the quadrupole polynomials $\{f_n^\xi\}$. That is,

$$\int \rho^\xi(\gamma) f_n^\xi(\gamma) f_{n'}^\xi(\gamma) d\gamma = \delta_{nn'}. \qquad (2.7)$$

It should be obvious from Eq. (2.4) that the diagonal representation, where $H_\theta|\chi_n\rangle = E_\theta|\chi_n\rangle$, is obtained if and only if $\xi = 0$ and $\gamma = \pm\sqrt{4n^2 + 1/4}$. This means that a diagonal representation is obtained only in the absence of the electric quadrupole. This is the reason why the quadrupole potential in 2D, $\xi\frac{\sin 2\theta}{r^2}$, is not found in the literature among the elements of the class of exactly solvable problems. It is important to note that the diagonal representation we are referring to here is associated with the operator $H_\theta$. That is, $(H_\theta)_{nn'} = E_\theta \delta_{nn'}$. One should not confuse this with the discrete bound states

spectrum, which is associated with the diagonal representation of the total Hamiltonian $H$ (i.e., $H_{nn'} = E\delta_{nn'}$). Consequently, it is neither required nor necessary for bound states to have a vanishing quadrupole moment or that $\gamma$ be quantized as shown above. On the other hand, for bound states we take $2E_\theta = -\omega^2 - \frac{1}{4}$, where $\omega$ is a real dimensionless parameter. This is equivalent to the above with $\gamma \to i\omega$. Thus, the recursion relation (2.5) changes into

$$\tfrac{1}{4}\omega^2 f_n^\xi = -\left(n^2 + \tfrac{1}{16}\right) f_n^\xi + \xi\left(f_{n-1}^\xi + f_{n+1}^\xi\right). \tag{2.8}$$

We refer to the polynomial solution of this recursion relation as the "*quadrupole polynomials*" of the second kind. They are completely defined by this recursion and the initial values $f_0^\xi = 1$ and $f_1^\xi = \frac{1}{4\xi}\left(\frac{1}{4} + \omega^2\right)$. One can easily show that the Chebyshev polynomial of the second kind is a special limiting case of this polynomial as follows

$$U_n(x) = \lim_{|\xi| \to \infty} f_n^\xi\left(2\sqrt{2\xi x}\right). \tag{2.9}$$

The polynomial solution of (2.5) are the "*quadrupole polynomials*" of the first kind. By comparing the recursion (2.5) with (2.8) we can relate these two kinds by the parameter map $\gamma \to i\omega$. Therefore, the bound states angular component of the wavefunction is identical to (2.6) but with $\gamma \to i\omega$. However, relation (2.8) implies that if $\xi$ becomes too small then reality of the representation will be compromised. To maintain reality, the quadrupole moment should be greater than a certain critical value. To obtain this critical quadrupole value, we investigate Eq. (2.8), which could be written as the eigenvalue equation $h|f\rangle = \frac{1}{4}\omega^2|f\rangle$, where $h$ is the tridiagonal symmetric matrix

$$h_{nm} = -\left(n^2 + \tfrac{1}{16}\right)\delta_{nm} + \xi\left(\delta_{n,m+1} + \delta_{n,m-1}\right). \tag{2.10}$$

Therefore, the determinant of the matrix $h - \frac{1}{4}\omega^2 I$ must vanish, where $I$ is the identity matrix. This requirement is necessary so that the kernel of this matrix operator becomes non-singular and prevents the solution space from being empty. Consequently, this translates into a condition on the electric quadrupole moment $\xi$ that depends on $\omega$. The critical value of $\xi$ is the smallest value that satisfies this condition. It corresponds to the case where the eigenvalue of the matrix $h$ vanishes (i.e. when $\omega = 0$). Table 1 shows a sequence of these critical values for an $N$-dimensional matrix $h$ with $N = 2, 3, .., 7$. It is evident that the sequence converges rapidly with $N$ for the given choice of significant digits. In fact, for each $N$ one finds a set of $2K$ zeros, $\{\pm\xi_i\}_{i=1}^K$, of the determinant where $K = \frac{N}{2}$ or $K = \frac{N-1}{2}$ if $N$ is even or odd, respectively. The positive zeros (for large enough $N$) are the critical values of the quadrupole moment corresponding to all vibrational states (the ground state as well as exited states). Table 2 lists the lowest positive zeros for large $N$.

## 3. The radial component of the wavefunction

The radial wavefunction $R(r)$ could be taken as an element in the space spanned by the following $L^2$ functions that are compatible with the domain of the radial component of the Hamiltonian $H_r$:

$$\phi_k(y) = b_k y^\beta e^{-y/2} L_k^\alpha(y), \tag{3.1}$$

where $y = \lambda r$, $k = 0,1,2,...$, and $L_k^\alpha(y)$ is the associated Laguerre polynomial of degree $k$. The real parameter $\lambda$ is positive with an inverse length dimension (i.e., it is a length scale parameter). On the other hand, the dimensionless parameters $\beta > 0$ and $\alpha > -1$. The normalization constant, $b_k = \sqrt{\lambda \Gamma(k+1)/\Gamma(k+\alpha+1)}$, is chosen to conform with the orthogonality property of the Laguerre polynomials. Using the differential equation (B.3) and differential formula (B.4) for the Laguerre polynomials, we can calculate the action of the radial differential wave operator in Eq. (1.4b) on the basis element (3.1) giving

$$(H_r - E)|\phi_k\rangle = \frac{\lambda^2}{2}\left[\frac{k}{y}\left(1 + \frac{\alpha+1-2\beta}{y}\right) + \frac{\gamma^2 - (\beta-\frac{1}{2})^2}{y^2} + \frac{\beta}{y} - \frac{1}{4} - \frac{2}{\lambda^2}E\right]$$
$$+ \frac{\lambda^2}{2} \frac{(k+\alpha)(2\beta-\alpha-1)}{y^2} \frac{b_k}{b_{k-1}}|\phi_{k-1}\rangle \quad (3.2)$$

Here, we are considering the *continuum scattering states* ($E > 0$) where we took $2E_\theta \equiv \gamma^2 - \frac{1}{4}$. The recurrence relation (B.1) and orthogonality formula (B.5) for the Laguerre polynomials show that a tridiagonal matrix representation $\langle\phi_k|H_r - E|\phi_{k'}\rangle$ is possible for positive energies if and only if $\alpha = 2\beta - 1$ and $(\beta - \frac{1}{2})^2 = \gamma^2$. That is,

$$\beta = \begin{cases} \frac{1}{2} + \gamma & ; \gamma > 0 \\ \frac{1}{2} - \gamma & ; \gamma < 0 \end{cases} \quad (3.3)$$

which makes $\beta$ always greater than or equal to $+\frac{1}{2}$ and $\alpha = \pm 2\gamma$ for $\pm\gamma > 0$. We expand the radial component of the wavefunction in the basis (3.1) as $R(r) = \sum_{k=0}^{\infty} g_k^\gamma(E)\phi_k(y)$ and substitute this in the wave equation (1.4b). Employing the action of the wave operator on the basis as given by Eq. (3.2) and using the recurrence relation and orthogonality formula of the Laguerre polynomials, we obtain the following tridiagonal matrix representation of the radial wave operator

$$\langle\phi_k|H_r - E|\phi_{k'}\rangle = (2k+\alpha+1)\left(\frac{\lambda^2}{8} - E\right)\delta_{k,k'}$$
$$+ \left(\frac{\lambda^2}{8} + E\right)\left[\sqrt{k(k+\alpha)}\delta_{k,k'+1} + \sqrt{(k+1)(k+\alpha+1)}\delta_{k,k'-1}\right] \quad (3.4)$$

Therefore, the resulting three-term recursion relation for the expansion coefficients of the radial wavefunction becomes

$$(2k+\alpha+1)(\cos\varphi)g_k^\gamma = \sqrt{k(k+\alpha)}\, g_{k-1}^\gamma + \sqrt{(k+1)(k+\alpha+1)}\, g_{k+1}^\gamma, \quad (3.5)$$

where $\cos\varphi = \frac{8E-\lambda^2}{8E+\lambda^2}$. Rewriting this recursion in terms of the polynomials $S_k^\gamma(E) = b_k g_k^\gamma(E)$, we obtain a familiar recursion relation as follows

$$2\left(k + \frac{\alpha+1}{2}\right)(\cos\varphi)S_k^\gamma = k\, S_{k-1}^\gamma + (k+\alpha+1)S_{k+1}^\gamma. \quad (3.6)$$

We compare this three-term recursion relation to that of the IUSG polynomial (B.6) in Appendix B. Thus, for the continuum case, we obtain the following $L^2$–series solutions for the radial component of the wavefunction

$$R(r) = B_\gamma^E (\lambda r)^{\frac{\alpha+1}{2}} e^{-\lambda r/2} \sum_{k=0}^{\infty} D_k^{\frac{\alpha+1}{2}}(\cos\varphi)L_k^\alpha(\lambda r), \quad (3.7)$$

where $\alpha = \pm 2\gamma$ for $\pm\gamma > 0$. The normalization constant $B_\gamma^E$ depends on $\gamma$ and the energy but, otherwise, independent of $r$ and $k$. To make $R(r)$ energy-normalized we write $B_\gamma^E = (\sin\varphi)^{1+\frac{\alpha}{2}}/\sqrt{E}$.

The basis (3.1) is, sometimes, referred to as the "Laguerre basis". There exist another equivalent representation for the continuum radial wavefunction in a space spanned by the same basis but with $y = (\lambda r)^2$ and $b_k = \sqrt{2\lambda \Gamma(k+1)/\Gamma(k+\alpha+1)}$. This basis is referred to as the "oscillator basis". A tridiagonal matrix for the radial wave operator in this basis is also possible but only if $2\beta = \alpha + \frac{1}{2}$ and $\alpha = \pm\gamma$ for $\pm\gamma > 0$. Writing $g_k^\gamma(E) = b_k S_k^\gamma(E)$, the resulting three-term recursion relation, in this case, reads as follows

$$\frac{2E}{\lambda^2} S_k^\gamma = (2k+\alpha+1) S_k^\gamma + (k+\alpha) S_{k-1}^\gamma + (k+1) S_{k+1}^\gamma, \tag{3.8}$$

Comparing this with the recursion relation (B.1) for the Laguerre polynomials, we conclude that $S_k^\gamma(E) \propto (-1)^k L_k^\alpha(2E/\lambda^2)$. Thus, we can write the following radial component of the continuum wavefunction

$$R(r) = B_\gamma^E (\lambda r)^{\alpha+\frac{1}{2}} e^{-\lambda^2 r^2/2} \sum_{k=0}^{\infty} (-1)^k \frac{\Gamma(k+1)}{\Gamma(k+\alpha+1)} L_k^\alpha(2E/\lambda^2) L_k^\alpha(\lambda^2 r^2). \tag{3.9}$$

The sum in this expression could be written in a closed form [10]. If we also impose $L^2$ energy normalization by using the weight function for the Laguerre polynomials $L_k^\alpha(2E/\lambda^2)$, we finally obtain

$$R(r) = \frac{B_\alpha}{2^{\alpha+1}\Gamma(\alpha+1)} \left(\tfrac{1}{2} E r^2\right)^{\frac{\alpha}{2}+\frac{1}{4}} {}_0F_1\!\left(;\alpha+1;-\tfrac{1}{2} E r^2\right). \tag{3.10}$$

where $B_\alpha$ is an overall constant factor.

Now, we turn attention to the radial component of the wavefunction for *bound states*. In this case, we write $2E_\theta = -\frac{1}{4} - \omega^2$ and a real tridiagonal matrix representation $\langle \phi_k | H_r - E | \phi_{k'} \rangle$ is possible only in the Laguerre basis (3.1) with $2\beta = \alpha + 2$ and $\lambda^2 = -8E$. Thus, the basis becomes energy dependent via the parameter $\lambda$ and we obtain the following tridiagonal matrix elements for the radial wave operator in this basis

$$\langle \phi_k | H_r - E | \phi_{k'} \rangle = 4E\left[\omega^2 + \left(\tfrac{\alpha+1}{2}\right)^2 + k - (2k+\alpha+1)\left(k+\tfrac{\alpha}{2}+1\right)\right]\delta_{k,k'}$$
$$+ 4E\left[\left(k+\tfrac{\alpha}{2}\right)\sqrt{k(k+\alpha)}\,\delta_{k,k'+1} + \left(k+\tfrac{\alpha}{2}+1\right)\sqrt{(k+1)(k+\alpha+1)}\,\delta_{k,k'-1}\right] \tag{3.11}$$

Thus, the radial wave equation becomes equivalent to the following three-term recursion relation for the expansion coefficients of the wavefunction

$$\omega^2 S_k = \left[(k+\alpha+1)\left(k+\tfrac{\alpha}{2}+1\right) + k\left(k+\tfrac{\alpha}{2}\right) - \left(\tfrac{\alpha+1}{2}\right)^2\right] S_k$$
$$- k\left(k+\tfrac{\alpha}{2}\right) S_{k-1} - (k+\alpha+1)\left(k+\tfrac{\alpha}{2}+1\right) S_{k+1} \tag{3.12}$$

where $g_k^\omega(E) = b_k S_k(\omega;E)$. One should observe the curious absence of the energy from this recursion relation. This implies that the dependence of $S_k(\omega;E)$ on energy is via an overall factor, which is independent of $k$, say $F_\omega^\alpha(E)$. The source of this factorization comes from (3.11), where all elements of the matrix representation of the radial wave operator have the common factor $4E$. This means that the wave equation, in this representation, is satisfied independently of any value of the energy as long as it is negative (due to $\lambda^2 = -8E$). Consequently, this property has a dramatic implication on the bound states energy spectrum. It implies that for any choice of non-positive energy a bound state could be supported. However, the diagonal constraint on the representation (3.11) dictates that $E = 0$. These observations, concerning the peculiar behavior of the bound states energy spectrum, have already been reported and analyzed in the literature for the inverse square potential. See, for example, [7] and references therein. Regularization procedures [7,11] and self-adjoint extensions of the Hamiltonian [12] were introduced to handle these irregularities (anomalies).

Now, Eq. (3.12) is a special case of the three-term recursion relation of the continuous dual Hahn orthogonal polynomials, $Q_k^\mu(x;a,b)$, where $x$ is real and $\mu$, $a$, $b$ are positive except for a possible pair of complex conjugates with positive real parts [13]. The general recursion relation for these polynomials reads as follows

$$x^2 Q_k^\mu = \left[(k+\mu+a)(k+\mu+b) + k(k+a+b-1) - \mu^2\right] Q_k^\mu$$
$$-k(k+a+b-1)Q_{k-1}^\mu - (k+\mu+a)(k+\mu+b)Q_{k+1}^\mu \tag{3.13}$$

and with the standard initial seed, $Q_0^\mu = 0$. These polynomials could be written in terms of the generalized hypergeometric function as follows

$$Q_k^\mu(x;a,b) = {}_3F_2\left(\begin{matrix}-k,\mu+ix,\mu-ix\\ \mu+a,\mu+b\end{matrix}\bigg|1\right). \tag{3.14}$$

The associated orthogonality relation is

$$\int_0^\infty \rho^\mu(x) Q_n^\mu(x;a,b) Q_m^\mu(x;a,b)\, dx = \frac{\Gamma(n+1)\Gamma(n+a+b)}{\Gamma(n+\mu+a)\Gamma(n+\mu+b)}\delta_{nm}, \tag{3.15}$$

where the weight function is $\rho^\mu(x) = \frac{1}{2\pi}\left|\frac{\Gamma(\mu+ix)\Gamma(a+ix)\Gamma(b+ix)}{\Gamma(\mu+a)\Gamma(\mu+b)\Gamma(2ix)}\right|^2$. Comparing (3.13) with (3.12), we can write $g_k^\omega(E) \propto b_k\, Q_k^{\frac{\alpha+1}{2}}\left(\omega;\frac{\alpha+1}{2},\frac{1}{2}\right)$. Thus, the expansion coefficients are defined modulo the *arbitrary* factor $F_\omega^\alpha(E)$ that depends on $\omega$ and $E$. Finally, the radial component of the wavefunction for the bound state at energy $E$ becomes

$$R(E,r) = F_\omega^\alpha(E)(2\eta r)^{\frac{\alpha+1}{2}} e^{-\eta r} \sum_{k=0}^\infty \frac{\Gamma(k+1)}{\Gamma(k+\alpha+1)} Q_k^{\frac{\alpha+1}{2}}\left(\omega;\frac{\alpha+1}{2},\frac{1}{2}\right) L_k^\alpha(2\eta r), \tag{3.16}$$

where the energy dependent wave number $\eta$ is defined by $E = -\frac{1}{2}\eta^2$.

## 4. Conclusion and discussion

In a complete discrete $L^2$ basis, we removed the diagonal constraint on the matrix representation of the wave equation. That is, the energy eigenvalue equation is not restricted to the usual form $H|\phi_n\rangle = E_n|\phi_n\rangle$ but allowed to take the less constrained form $(H-E)|\phi_n\rangle \sim |\phi_n\rangle + |\phi_{n-1}\rangle + |\phi_{n+1}\rangle$. As a result, the solution space incorporates new problems that were not members of the class of exactly solvable potentials. In particular, recently we were able to obtain an exact solution for the noncentral dipole potential in 3D, $\cos\theta/r^2$ [4]. Thus, we were able to present an analytic description for the interaction of an electron with a molecule that has a permanent electric dipole moment. In the present work, we showed that it was also possible to extend the same development to the electric quadrupole potential in 2D, $\sin 2\theta/r^2$. However, our approach is not compatible with the electric dipole potential in 2D, $\sin\theta/r$, nor with the electric quadrupole potential in 3D.

In principle, the theoretical findings of this work could be tested in an appropriate experimental setup as was suggested in the introduction section. For example, one could verify the existence of a critical quadrupole moment (0.2557 a.u.) for an electron capture in a 2D system.

## Acknowledgments

I am grateful to H. Bahlouli for stimulating and fruitful discussions. The support provided by King Fahd University of Petroleum and Minerals is highly appreciated.

## Appendix A
### The Coulomb problem in any dimension, the n-dimensional Hydrogen atom, and the electric quadrupole potential in 2D

The Coulomb problem in any dimension is defined as that of an electrostatic point charge at the origin of free space. This problem in 3D is one of the most familiar of all problems in Physics. A simple method to visualize the Coulomb problem in *n* dimensions, where *n* is less than three, is as follows. In 3D, one constructs a uniform (3–*n*)-dimensional charge distribution having the desired *n*-dimensional symmetry. Consequently, one obtains the following situations in 1D, 2D, and 3D, respectively:

1) The Coulomb potential in one dimension with coordinate $x \in [-\infty, +\infty]$ is proportional to the distance from the origin (i.e., $V \sim |x|$) resulting in a constant electric field pointing everywhere away from, or towards, the charge. This could easily be understood by considering the problem of an infinite uniform surface charge distribution in the *y*-*z* plane. However, on the semi-infinite real line, $x \in [0, +\infty]$, $V \sim x$.

2) The Coulomb potential in two dimensions with circular polar coordinates $(r,\theta)$ is logarithmic (i.e., $V \sim \ln r$) resulting in an electric field which varies as $r^{-1}$. This could also be understood by considering the problem of an infinite uniform linear charge distribution along the *z*-axis.

3) The Coulomb potential in three dimensions with spherical polar coordinates $(r,\theta,\phi)$ is $V \sim r^{-1}$. This is the familiar 3D problem.

For $n \geq 3$, one can easily deduce that the Coulomb potential in *n*-dimensional Euclidean space is $V(r) \sim r^{2-n}$, where $r = \sqrt{x_1^2 + x_2^2 + ... + x_n^2}$ is the distance from the point charge. Now, the problem in *n* dimensions with $V(r) \sim r^{-1}$ has sometimes been referred to in the literature by the more appropriate term: the "*n*-dimensional Hydrogen atom" [14]. It is only for *n* = 3 where this and the Coulomb problem coincide. Now, to the 2D problem.

We represent the 2D Coulomb problem in three dimensions by an infinite linear charge of uniform density $\sigma$ (charge per unit length) along the *z*-axis. It is an elementary exercise to find the electrostatic potential due to this charge distribution in the 2D *xy*-plane (equivalently, the 2D cylindrical $r\theta$-plane) as

$$V(r) = V(r_0) - \frac{\sigma}{2\pi\epsilon_0}\ln(r/r_0),\tag{A.1}$$

where $r_0$ is a reference radius. Now, going back to the 2D problem and writing the value of the charge at the origin as $Q = \sigma r_0$, the Coulomb potential in 2D for a point charge *Q* (monopole) at the origin is written as

$$V_M(r) = V(r_0) - \frac{Q}{2\pi\epsilon_0 r_0}\ln(r/r_0).\tag{A.2}$$

Using this result, one can find the potential of the electric dipole in 2D shown in Fig. 1 with $\frac{d}{r} \to 0$ as $V_D(r,\theta) = V_M(r_+) - V_M(r_-)$, where *d* is the distance between the two charges and $r_\pm = r \mp \frac{1}{2}d\sin\theta$. The result is

$$V_D(r,\theta) = \frac{p_D}{2\pi\epsilon_0 r_0}\frac{\sin\theta}{r},\tag{A.3}$$

where the electric dipole moment $p_D = Qd$. Similarly, using this result, one could find the potential of the electric quadrupole configuration shown in Fig. 2 with $\frac{d}{r} \to 0$. This could be done (as shown in Fig. 3) by writing $V_Q(r,\theta) = V_D(r_+,\theta_+) - V_D(r_-,\theta_-)$, where now $r_\pm = r \mp \frac{1}{2}d\cos\theta$ and $\theta_\pm = \theta \pm \frac{d\sin\theta}{2r}$ giving

$$V_Q(r,\theta) = \frac{p_Q}{2\pi\epsilon_0 r_0}\frac{\sin 2\theta}{r^2},\tag{A.4}$$

where the electric quadrupole moment $p_Q = Qd^2$. However, if the quadrupole were to be constructed from a charge $+2Q$ at the origin and two $-Q$ charges located at $(r,\theta) = (d, \pm\frac{\pi}{2})$ then the potential will be the same as (A.4) except that $\sin 2\theta$ is replaced by $\cos 2\theta$. Now in the limit as $d \to 0$, the length scale needed to define $r_0$ is lost. However, if a test particle of mass *m* and charge *q* is introduced in the problem, then we recover the length scale as, for example, $p_D/q$, $\sqrt{p_Q/q}$, or $a_0 = 4\pi\epsilon_0\hbar^2/mq^2$. Note that the Compton wavelength $\frac{\hbar}{mc}$ does not come into play since the problem is nonrelativistic ($c \to \infty$). Now, in the monopole case we can only use $a_0$ as the length scale. In this work, our choice is $r_0 = \frac{1}{2}a_0$. Of course, one can make a different choice for the factor that multiplies $a_0$ by simply redefining (rescaling) the charge *Q* or moments.

## Appendix B
### The associated Laguerre polynomials and the *improved* ultra-spherical (Gegenbauer) polynomials

(1) The associated Laguerre polynomials $L_n^\nu(x)$, where $x \in [0, +\infty]$ and $\nu > -1$:

$$xL_n^\nu(x) = (2n+\nu+1)L_n^\nu(x) - (n+\nu)L_{n-1}^\nu(x) - (n+1)L_{n+1}^\nu(x) \tag{B.1}$$

$$L_n^\nu(x) = \frac{\Gamma(n+\nu+1)}{\Gamma(n+1)\Gamma(\nu+1)}\,{}_1F_1(-n;\nu+1;x) \tag{B.2}$$

$$\left[x\frac{d^2}{dx^2} + (\nu+1-x)\frac{d}{dx} + n\right]L_n^\nu(x) = 0 \tag{B.3}$$

$$x\frac{d}{dx}L_n^\nu = nL_n^\nu - (n+\nu)L_{n-1}^\nu \tag{B.4}$$

$$\int_0^\infty x^\nu e^{-x} L_n^\nu(x) L_m^\nu(x)\,dx = \frac{\Gamma(n+\nu+1)}{\Gamma(n+1)}\delta_{nm} \tag{B.5}$$

(2) The *improved* ultra-spherical (Gegenbauer) polynomials $D_n^\mu(x)$, where $x \in [-1,+1]$ and $\mu > -\tfrac{1}{2}$ [9]:

$$2(n+\mu)\,x\,D_n^\mu(x) = n\,D_{n-1}^\mu(x) + (n+2\mu)\,D_{n+1}^\mu(x) \tag{B.6}$$

$$D_n^\mu(x) = \frac{\Gamma(\mu+1/2)\Gamma(n+1)}{\Gamma(n+\mu+1/2)} P_n^{(\mu-\tfrac{1}{2},\mu-\tfrac{1}{2})}(x) = {}_2F_1\left(-n, n+2\mu; \mu+\tfrac{1}{2}; \tfrac{1-x}{2}\right) \tag{B.7}$$

$$\left[(1-x^2)\frac{d^2}{dx^2} - (2\mu+1)x\frac{d}{dx} + n(n+2\mu)\right]D_n^\mu(x) = 0 \tag{B.8}$$

$$(1-x^2)\frac{d}{dx}D_n^\mu = -n\,x\,D_n^\mu + n\,D_{n-1}^\mu \tag{B.9}$$

$$\int_{-1}^{+1}(1-x^2)^{\mu-\tfrac{1}{2}} D_n^\mu(x) D_m^\mu(x)\,dx = 2^{2\mu-1}\frac{\Gamma(\mu+1/2)^2\,\Gamma(n+1)}{(n+\mu)\Gamma(n+2\mu)}\delta_{nm} \tag{B.10}$$

course, with the same weight function, roots, etc.) because it does not suffer from this ill-definition at $\mu = 0$. For the case $n = \mu = 0$, the right-hand-side of the orthogonality relation (B.10), when evaluated using $\lim_{z \to 0}[z\Gamma(z)] = 1$, gives the value $\pi$.

**Tables caption**

**Table 1:** A sequence of the lowest eigenvalue of the $N \times N$ tridiagonal matrix (2.10) converging with $N$ to the critical value of the quadrupole moment (in atomic units).

**Table 2:** The smallest positive eigenvalues of the $N \times N$ tridiagonal matrix (2.10) for $N = 200$. They correspond to the critical values of the quadrupole moment (in atomic units) for the ground state ($n = 0$) as well as few of the lowest exited states ($n \geq 1$).

**Table 1**

| $N$ | $\xi$ (a.u.) |
|---|---|
| 2 | 0.257694101601104 |
| 3 | 0.255734422198048 |
| 4 | 0.255730974701113 |
| 5 | 0.255730973148881 |
| 6 | 0.255730973148629 |
| 7 | 0.255730973148629 |

**Table 2**

| $n$ | $\xi$ (a.u.) |
|---|---|
| 0 | 0.2557309731486 |
| 1 | 5.4091751672900 |
| 2 | 17.434377594381 |
| 3 | 36.338459922255 |
| 4 | 62.118980707609 |
| 5 | 94.775228357085 |
| 6 | 134.30694411901 |
| 7 | 180.71401152335 |
| 8 | 233.99637041731 |
| 9 | 294.15398660063 |
| 10 | 361.18683919318 |

**Figures caption**

**Fig. 1:** The electric dipole configuration in 2D with $d \ll r$.

**Fig. 2:** An electric quadrupole configuration in 2D with $d \ll r$.

**Fig. 3:** The electric quadrupole is the sum of two electric dipoles.

**Fig. 4:** The 2D electric quadrupole of Fig. 2 is equivalent, in 3D, to four thin parallel wires with uniform charge distribution. Each wire is at a distance $d$ from its nearest neighbors. The charge $q$ moves in a plane normal to the wires.

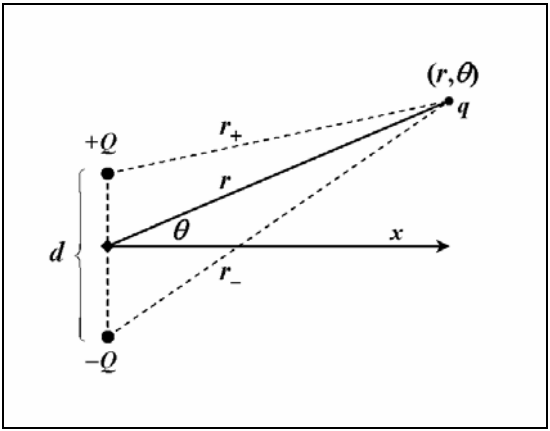

**Fig. 1**

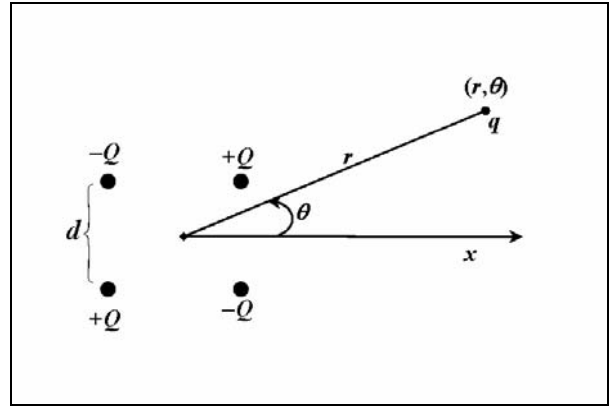

**Fig. 2**

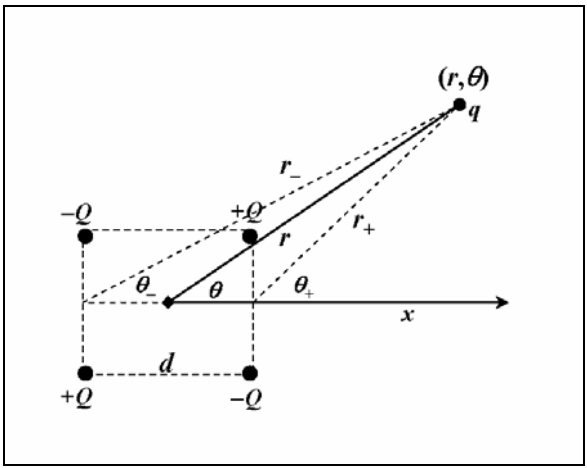

**Fig. 3**

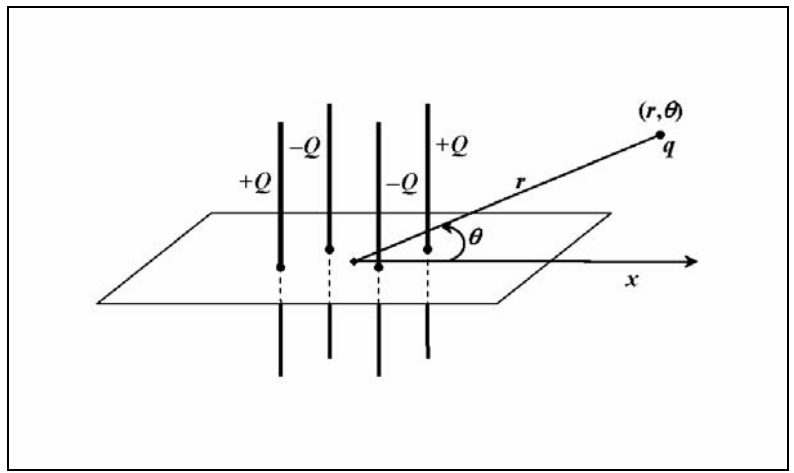

**Fig. 4**